\documentclass[lnicst,10pt]{svmultln}

\usepackage{makeidx}  
\usepackage{url}
\usepackage{graphicx}
\usepackage{color}
\usepackage[T1]{fontenc}
\usepackage{longtable}
\usepackage{tabularx}
\usepackage{array}
\usepackage{amsmath}
\usepackage{hyperref}
\usepackage{cleveref}
\usepackage{multirow}
\usepackage{color}
\usepackage{enumitem}
\usepackage[table]{xcolor}
\usepackage[
  pass,
]{geometry}

\definecolor{applegreen}{rgb}{0.55, 0.71, 0.0}

\begin{document}
\mainmatter              
\title{Poor Peering: a reflexion about a RIXP}
\titlerunning{RIXP IOA}  
%
\author{Rehan Noordally \inst{1}\and Xavier Nicolay\inst{1} \and Pascal Anelli \inst{1}\and Richard Lorion \inst{2}\and Tahiry Razafindralambo\inst{1}}
\authorrunning{Rehan Noordally et al.}   
%
\tocauthor{Rehan Noordally, Xavier Nicolay, Pascal Anelli, Richard Lorion, Tahiry Razafindralambo}
\institute{Laboratoire d'Informatique et de Math\'ematiques, \\
\and
Laboratoire d'Energ\'etique, d'Electronique et Proc\'ed\'es,\\
University of Reunion Island,\\
15 Rue Ren\'e Cassin,\\
97490 Sainte Clotilde,\\
R\'eunion\\
email: firstname.lastname@univ-reunion.fr}

\maketitle              

\begin{abstract}        
Since more than twenty years, Internet evolves as well as users' needs. and the users' needs too. The need of bandwidth growths every day with the new usages. In this context, with megabit per second count. There is a real business behind our free access to web site. From a group of island, the economy of bandwidth can be a  major challenge. Installing a Regional Internet eXchange Point (RIXP) can be a solution to keep regional Internet traffic.
This paper brings a reflexion about where to install a RIXP in the Indian Ocean Area (IOA).
As there is a lot of different countries and stakes, we (stay) only on an scientist approach.

\keywords {Metrology, Active measurement, End-to-end delay, Peering, Route, Network, IXP.}
\end{abstract}

\section{Introduction}


The Internet was first realized to delivered information in a limited time with limited user interaction, like sending email or using the \emph{World Wide Web (WWW)}. With time, the use of the Internet have evolves and the user experience too. Now we currently use Internet to practice on-line gaming, audio or video discussion. These new services requires low latency for interactivity.

However, Internet are not fairly distribute all aver the world. Some region like the Indian Ocean Area (IOA) have poorly meshed topology. This none diversity of route impact the performance of a transport protocol, like \emph{Transmission Control Protocol (TCP)}. 



	In \cite{Nicolay2017-2}, Nicolay \textit{et Al.} shown that in IOA, the delay and the part are longer than necessary. It is worse when the data need to stay in the region. By reading the different official \emph{Internet eXchange Point (IXP)} websites, 
we learn that the regional peering are non-existent. With the analysis of data present in \cite{Nicolay2017-2}, they prove that the peering are made specifically in Europe. The increase of the delay and misrouting, the performance of TCP could be impacted. In \cite{Noordally2017}, Noordally \textit{et. Al} shown that is not the case for their laboratory based on Reunion Island. Their work was not representative of the Island's traffic but could give an idea and the major part of the measured traffic goes directly to Europa and North America.\\

 



As IOA peering is non-existing, we encourage the creation of a \emph{Regional Internet eXchange Point (RIXP)}. This material could, with the help of the \emph{Internet Services Provider (ISP)}, optimize the regional routes. This research and therefore proposals, based on scientists purposes, have been led free of geopolitic or economic considerations.\\

Our main contribution is to analyze the delay information from the IOA to identify the best location for a RIXP.\\ 

The remainder of this paper is organized as follows. Section \ref{sec:description} describes the topology of the submarines cables connecting Indian Ocean's Island to the Internet as well as the \emph{Internet eXchange Point (IXP)}. Section \ref{sec:measurements} presents our measurement setup deployed to decided the location of the RIXP. The results are analyzed in Section \ref{sec:results}. Section \ref{sec:related} reviews the related work. Finally, we conclude in section \ref{sec:conclusion}.

\section{Background}
\label{sec:description}

\begin{figure}[ht!]
\centering
\includegraphics[width=0.7\textwidth]{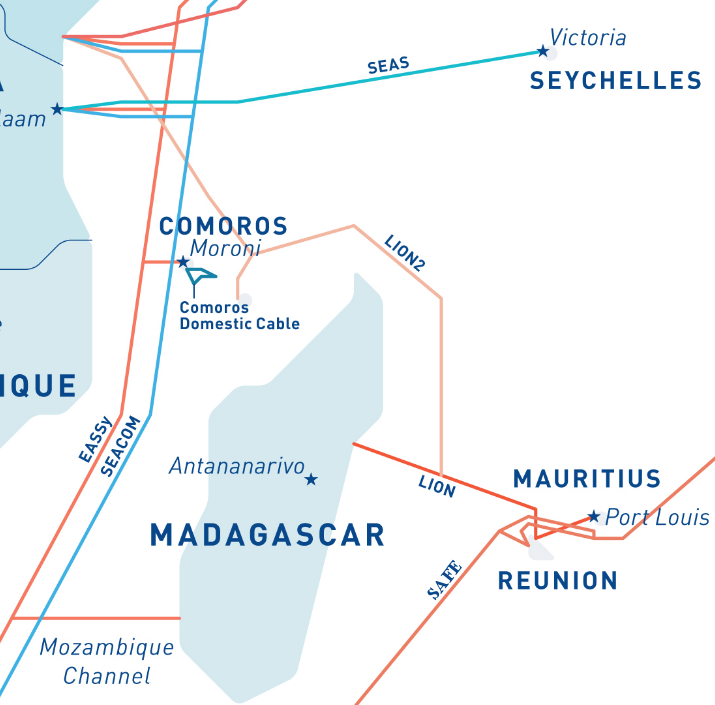}
\caption{Mascarene islands submarine cable \cite{cablesmap}}
\label{map:cable}
\end{figure}

We define the (IOA) as: Madagascar (MG), Mauritius (MU), Mayotte (YT), Reunion Island (RE) and Seychelles (SC). The map in Figure~\ref{map:cable} shows that each Island is connected to the Internet with one or more submarine cables. We can notice that LION/LION2 cable provides a link between Mayotte, Madagascar, Reunion and Mauritius. Each of these 4 islands have also an IXP. This equipment can be used by each ISP connected to it to exchange their traffic.\\

We know that:
   \begin{itemize}
      \item{ Each Country / Island are connected to one or more Submarines cables (see figure~\ref{map:cable}), so we know the real topology.}
      \item{ There is 4 IXP (Mayotte, Reunion, Mauritius, Madagascar). The AXIS Project~\cite{axis} have the goal to develop the installation of IXP in all African countries, but there is no obligation for the Internet Services Provider (ISP) to be connected with.}
      \item{ The are a few of ISP, none are present everywhere: ComoresTelecom, Emtel, CEB FiberNET Co Ltd, Blueline, Telma, Canal + Telecom, Orange, SRR (SFR R\'eunion), STOI, Telco OI (Only), Zeop, Airtel, Cable \& Wireless, Intelvision and Kokonet.}
   \end{itemize}
   
We do not know:
   \begin{itemize}
        \item {The logical path of a TCP/IP session. This information is related to the core objective of our paper. We aim at analyzing the Internet access of IOA islands }
        \item {The regional traffic in percent of international traffic. This information could help us in our analysis. Indeed, Internet access performance is strongly correlated with traffic shapes}
        \item {The capacity of each Internet Service Provider. This information could provide us some intuition on the traffic and peering policy of each ISP.}
	\end{itemize}

In this paper, based on our knowledge of the IOA islands Internet architecture, we aim at purpose to the impact of the installation of a RIXP.

\section{Measurement operations}
\label{sec:measurements}

We study the impact of the installation of a RIXP in the IOA. The country will be selection based on the theoretical delay. In the literature, different formula existing about the relationship between geographical distance and delay.\\

At first,  we need to identify the distance between each islands. For this calculation, we used the main city of each countries, and more especially the contact details of the city hall. With the address and the web service Google Map \cite{gmaps},  we obtain the latitude and longitude of each city hall. This information was necessary for the calculation of the distance between two countries. The Equation \ref{eq:distance} referrers to the distance between two point on The Earth. 
\begin{align}
   \begin{split}
      d & = arccos[cos(x) \times cos(y) \times cos(m) \times cos(n) \ + \\
        & +\ cos(x) \times sin(y) \times cos(m) \times sin(n) \ + \\
        & +\ sin(x)  \times sin(m)] \times 6371.1\ [km]	
   \end{split}
   \label{eq:distance}
\end{align}
In this formula, \{x,y\} and \{m,n\} are the geographical coordinates for two different city hall in two different countries. The tabular \ref{tab:distance} shows all distance obtained by the equation \ref{eq:distance}.\\

\begin{table}[ht!]
\centering
\begin{tabular}{c|c|c|c|c|c|}
\cline{2-6}
& MG & MU & RE & SC & YT \\ \hline
\multicolumn{1}{|c|}{MG} & 0 & $1,055.11$ & $857.09$ & $1,806.15$ & $723.75$ \\ \hline
\multicolumn{1}{|c|}{MU} & $1,055.11$ & 0 & $228.37$ & $1,742.28$ & $1,543.70$ \\ \hline
\multicolumn{1}{|c|}{RE} & $857.019$ & $228.37$ & 0 & $1,807.50$ & $1,410.83$ \\ \hline
\multicolumn{1}{|c|}{SC} & $1,806.15$ & $1,742.28$ & $1,807.50$ & 0 & $1,442.87$ \\ \hline
\multicolumn{1}{|c|}{YT} & $723.75$ & $1,543.70$ & $1,410.83$ & $1,442.87$ & 0 \\ \hline
\end{tabular}
\caption{Distance (in Km) between the different city hall of the capital city of the IOA}
\label{tab:distance}
\end{table}

With $228.37\ Km$ away, Mauritius and Reunion have the shortest distance. The maximal distance do not exceed $1,807.50\ km$, between Reunion Island and Seychelles.\\

The first formula about distance and delay learn by every one is the speed's equation. In \cite{Snyder2012}, the authors present the equation \ref{eq:speed}.

\begin{equation}
RTT = 2 \times \frac{distance}{\frac{2}{3} speed}
\label{eq:speed}
\end{equation}

The previous formula is based on the formula of the speed light in the vacuum. The propagation time needed to be updated with the inverse refractive index for the fiber. The multiplication by 2 is obligatory to get the \emph{Round-Trip Time (RTT)}.  

In \cite{Nicolay2017-2}, we have calculate the impact of the distance on the delay. The data used are coming from a measurement studies from IOA to world. From these data, we extract two formulas.
\begin{equation}
RTT = -0.00340391 \times distance + 431.557
\label{eq:ioa_world}
\end{equation}
\begin{equation}
RTT = 0.034018 \times distance + 328.092
\label{eq:ioa_ioa}
\end{equation}
The equation \ref{eq:ioa_world} group destination distributed all over the world, when the equation \ref{eq:ioa_ioa} concern only IOA's destination.

In \cite{Krajsa2011}, the author studies the relationship between delay and geographical distance. With a active metrology campaign, Krajsa \emph{ et al.} prove that, in certain condition, the delay could be approximate with the equation \ref{eq:krajsa}

\begin{equation}
RTT = 0.00128 * distance
\label{eq:krajsa}
\end{equation}

\subsubsection*{limits of the equations}
The equation \ref{eq:speed} is not adapted in the IOA's situation. It does not take into consideration the emission, queuing and treatment time. 

The equation \ref{eq:ioa_world} are decrease with distance. In \cite{Nicolay2017-2}, we have shown that the peering of IOA countries are made in Europe. This is explain the minus sign in front of the equation. 

The equation \ref{eq:ioa_ioa} is the actual delay we have measured. We need to reduce it by peering in the IOA. With this new routing rules, we can remove the 
\textit{y-intercept}.

The equation \ref{eq:krajsa} have been obtained in well-meshed area. In the IOA situation, we need to bring the actual delay to the delay we will obtained by this equation.\\

With this information, we will used only the two last equations presented.

\section{Results}
\label{sec:results}
As explain in the previous section, we will only used the equations \ref{eq:ioa_ioa} and \ref{eq:krajsa}. We will study the results in ISO-3166 country code \footnote{\href{https://fr.wikipedia.org/wiki/ISO\_3166}{https://fr.wikipedia.org/wiki/ISO\_3166}} order.\\


The following heat-map have been split in two part. The separation was represent by the diagonal in black. We considered that each countries have its own IXP and the delay to reach a local destination will be equal to 0. At the bottom, we have the delay obtain by \ref{eq:krajsa} and on the top the results obtain by \ref{eq:ioa_ioa}. 
The distance used by the two formula consist on the addition of the distance between two peer of countries. For example, if the RIXP are installed in Madagascar, the distance \{Mauritius$\rightarrow$Reunion Island\} could be factorize in \{Mauritius$\rightarrow$Madagascar\} + \{Madagascar$\rightarrow$Reunion Island\}.

\subsection*{Madagascar's case}





\begin{figure}[ht!]
\centering
\includegraphics[width=0.98\textwidth]{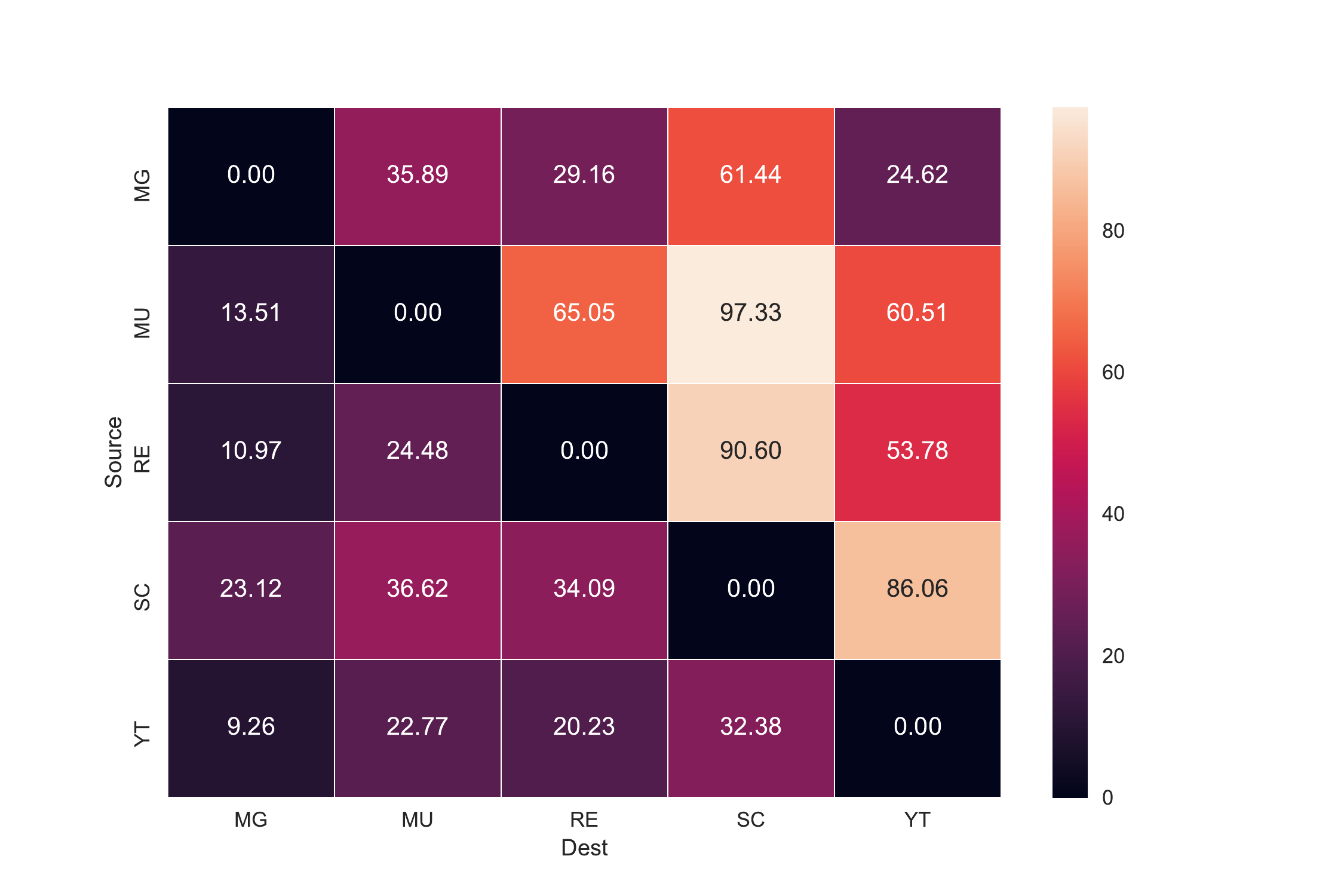}
\caption{RIXP in Madagascar Latency heatmap}
\label{curves:rixp_mg}
\end{figure}

The figure \ref{curves:rixp_mg} shows the delay between each IOA countries, if a RIXP are installed in Madagascar.\\ 

We can see that the delay on the top are far away to the delay obtain by the equation \ref{eq:krajsa}. With three bright cell, Madagascar have long delay when Seychelles and Mayotte are the source or the destination of the data.

\subsection*{Mauritius host the RIXP}





\begin{figure}[ht!]
\centering
\includegraphics[width=0.98\textwidth]{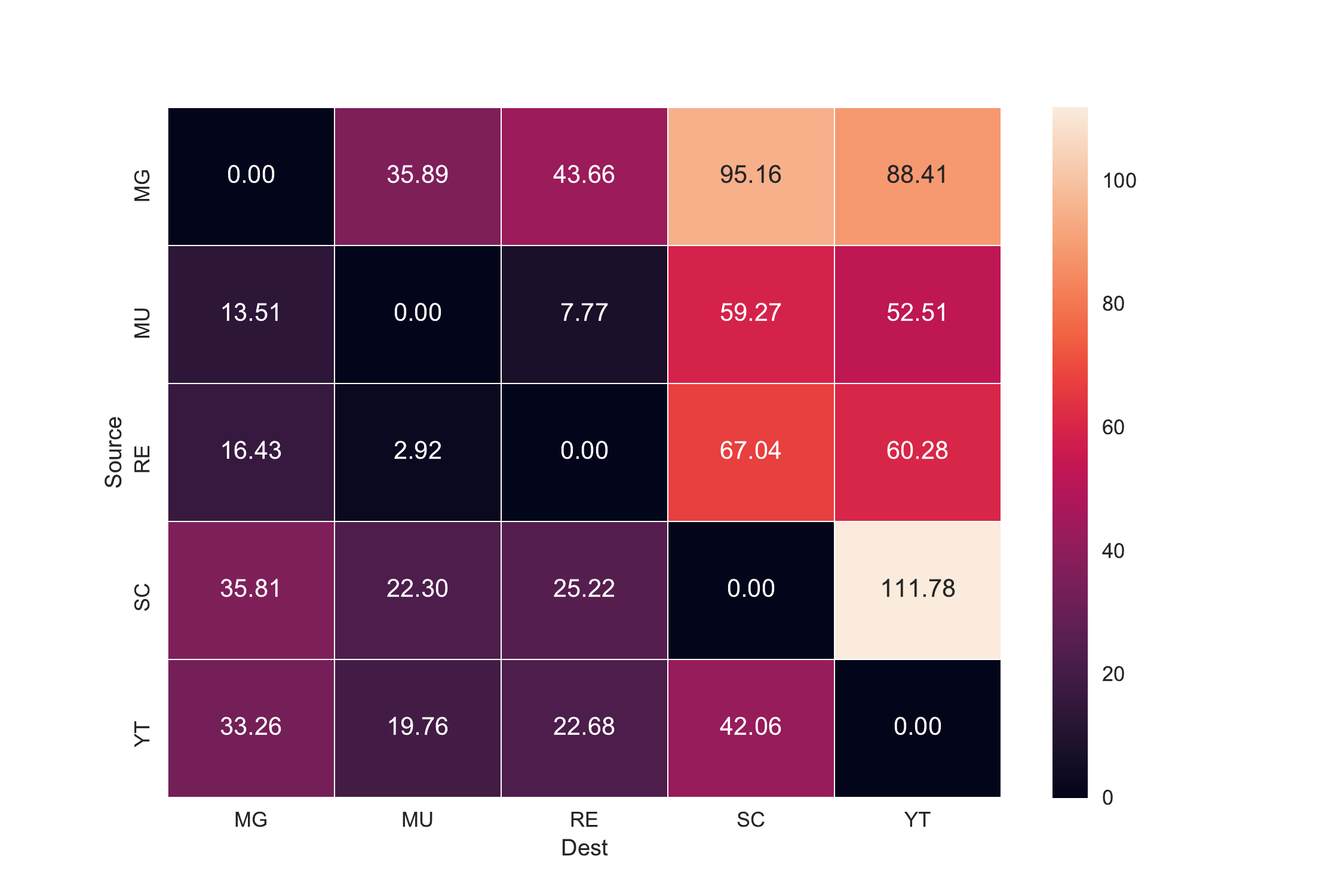}
\caption{RIXP in Mauritius Latency heatmap}
\label{curves:rixp_mu}
\end{figure}

The figure \ref{curves:rixp_mu} represent the delay obtain by the two equation when the RIXP are installed in Mauritius.\\

On the heat-map \ref{curves:rixp_mu}, the longer delay are obtain when Seychelles and Mayotte need to be exchange information. In the other case, the installation of the RIXP in Mauritius present advantages. Reunion Island and Madagascar will make profit with short delay when they will exchange between them but also with Mayotte.

\subsection*{Case of RIXP installed in Reunion Island}





\begin{figure}[ht!]
\centering
\includegraphics[width=0.98\textwidth]{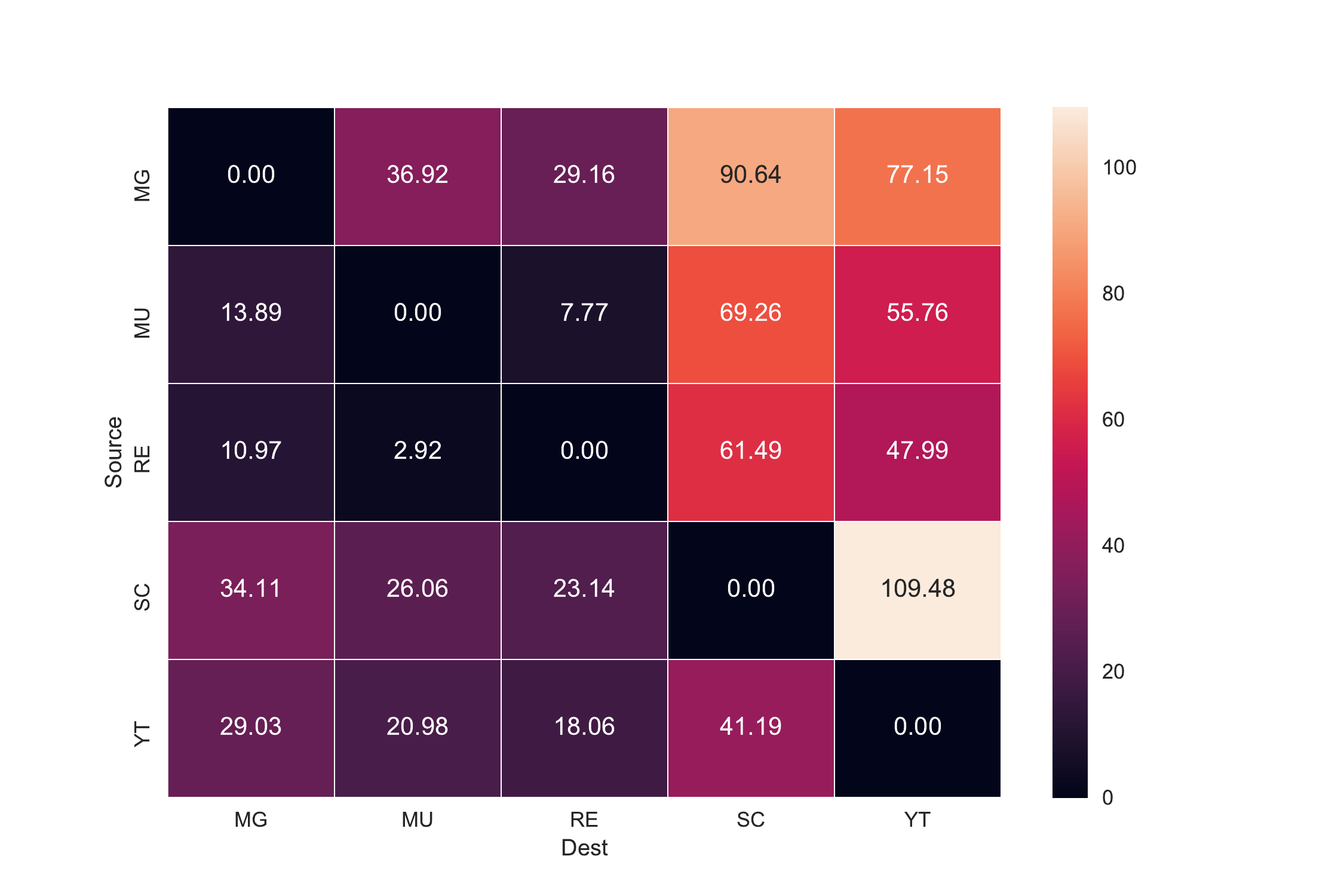}
\caption{RIXP in Reunion Latency heatmap}
\label{curves:rixp_re}
\end{figure}

The figure \ref{curves:rixp_re} illustrate the case of the installation of a RIXP in Reunion Island.\\

The analysis could be the same as the analysis from Mauritius. The main reason are the distance between the two island are minimal. But we can see that some delay are less than obtain by Mauritius case. 

\subsection*{The RIXP will be located in Seychelles}





\begin{figure}[ht!]
\centering
\includegraphics[width=0.98\textwidth]{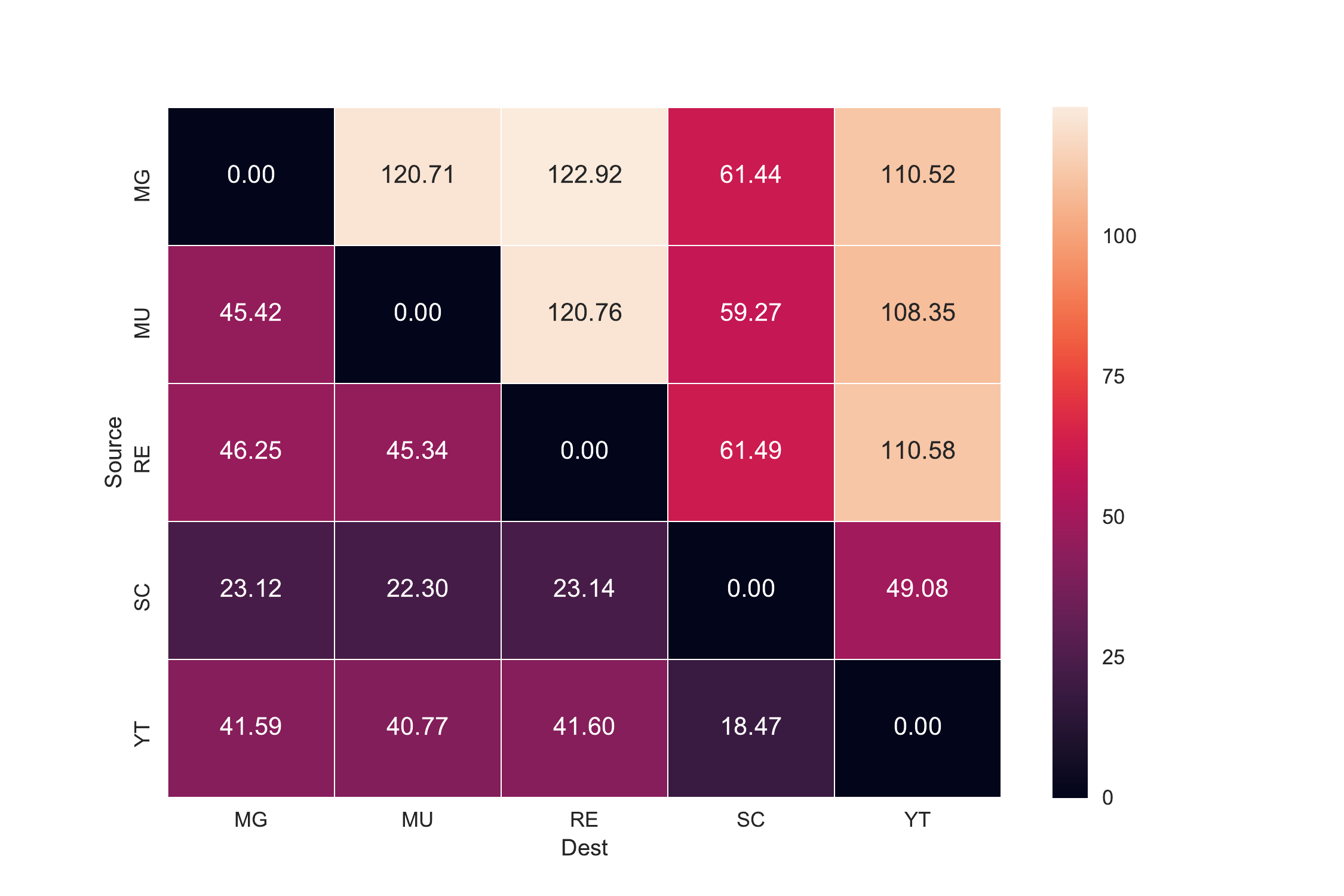}
\caption{RIXP in Seychelles Latency heatmap}
\label{curves:rixp_sc}
\end{figure}

The figure \ref{curves:rixp_sc} represent the delay obtain by the equations  \ref{eq:krajsa} and \ref{eq:ioa_ioa} if a RIXP are installed in Seychelles.

Seychelles are far away from the other islands. The delay obtained are longer than the case previously presented.

\subsection*{Mayotte study's case}


	


\begin{figure}[ht!]
\centering
\includegraphics[width=0.98\textwidth]{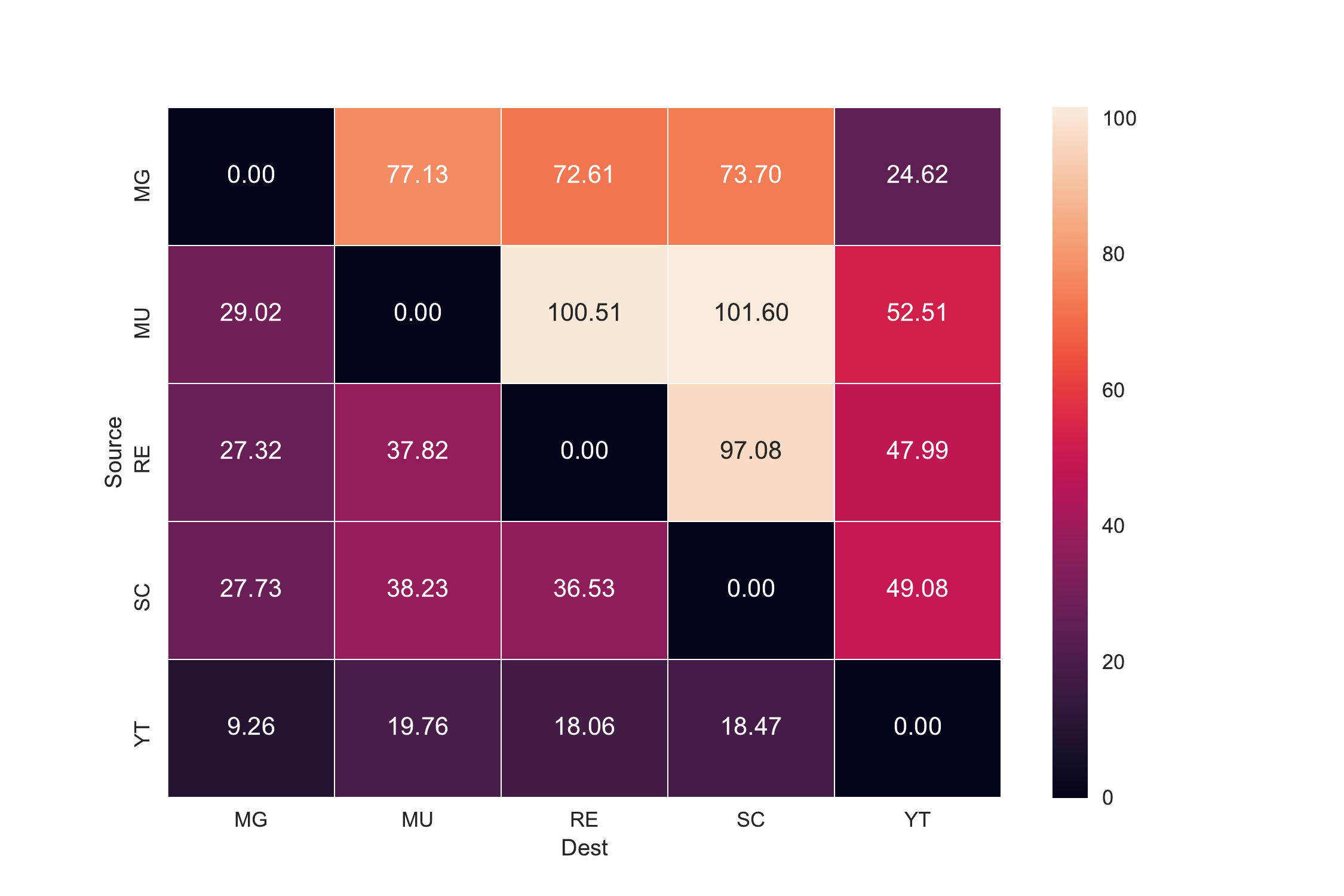}
\caption{RIXP in Mayotte Latency heatmap}
\label{curves:rixp_yt}
\end{figure}

The figure \ref{curves:rixp_yt} represent the Mayotte's case.

The number of bright cell in the figure \ref{curves:rixp_yt} are more important that the other case, except for Seychelles. This information mean the delay are longer than the other results.

\subsection{Comparison of results}




The previous heat-map represent the different possibility in term of delay dependent of the location of the RIXP. To keep in mind, more the cell are bright, more the delay is important.

First, we can remove the Seychelles and Mayotte as a possibility. When we count the cell in bright, the number are more important than in the other cases. Only stay, Madagascar, Mauritius and Reunion Island.

The first one have less dark cell than the two others. That mean the delay are longer than the two other possibilities. now we need, to compare cell by cell the results obtains by Mauritius and Reunion Island.

If the RIXP are installed in Reunion Island, the delay are, in majority, less than the delay obtain by a Mauritian RIXP.

With this results, only based on a delay analysis, we can purpose Reunion Island as the principal country for the installation of a RIXP.

\section{Related Work}
\label{sec:related}
The AXIS project is a project of the Internet Society  \cite{axis}. This project support the establishment of IXPs in African Union Member States. But this project does not make a reflexion about the interconnection of the countries.\\
In \cite{Soobron2014}, the authors have work on the installation of a Regional IXP in IOA. In this article, it is not the delay the main purpose. The choice was made on politics, economics and social criteria. The winner of this comparison was Mauritius Island.\\
In \cite{Galperin2013}, Galperin \& co works on the impact of the presence of a RIXP in Latin America and the Caribbean. They prove that the cost of inter-connections have been reduced. The installation of a RIXP could also reduced latency and enhanced quality of service. This point is only available if the server is located in a country connected to the RIXP.

\section{Conclusion}
\label{sec:conclusion}
Making a reflexion about the installation of a Regional Internet eXchange Point (RIXP) is a very important task in region where the delay are very high. The Indian Ocean Area (IOA) are connected to the Internet by only one or two submarines cables, depending of the country. From a previous study, we analyze the data produced and propose an enhancement for a better location for the installation of the equipment.\\
Our results shows that one country could reduce delay for most countries of the 
IOA.  This island is Reunion Island. We know that our results are not included some politics, economics and social criteria. The different government present in the IOA will fight for the installation of the materials in their lands.\\
The future step of our research concern the identification of the traffic exchange rate between the IOA countries. We leave for the future work the analysis of the TCP performance of the IOA, with the help of the different local ISP.
\bibliography{../../../metrology}

\begin{thebibliography}{1}
\providecommand{\url}[1]{#1}
\csname url@samestyle\endcsname
\providecommand{\newblock}{\relax}
\providecommand{\bibinfo}[2]{#2}
\providecommand{\BIBentrySTDinterwordspacing}{\spaceskip=0pt\relax}
\providecommand{\BIBentryALTinterwordstretchfactor}{4}
\providecommand{\BIBentryALTinterwordspacing}{\spaceskip=\fontdimen2\font plus
\BIBentryALTinterwordstretchfactor\fontdimen3\font minus
  \fontdimen4\font\relax}
\providecommand{\BIBforeignlanguage}[2]{{%
\expandafter\ifx\csname l@#1\endcsname\relax
\typeout{** WARNING: IEEEtran.bst: No hyphenation pattern has been}%
\typeout{** loaded for the language `#1'. Using the pattern for}%
\typeout{** the default language instead.}%
\else
\language=\csname l@#1\endcsname
\fi
#2}}
\providecommand{\BIBdecl}{\relax}
\BIBdecl

\bibitem{Nicolay2017-2}
X.~Nicolay, R.~Noordally, N.~M. Murad, and T.~Razafindralambo, ``Where is my
  next hop ? the case of indian ocean islands,'' in \emph{2017 Global
  Information Infrastructure and Networking Symposium (GIIS) (GIIS'17)}, St
  Denis, Reunion, Oct. 2017.

\bibitem{Noordally2017}
\BIBentryALTinterwordspacing
R.~Noordally, Y.~Gangat, A.~Ravoavahy, P.~Anelli, and X.~Nicolay, ``How long
  delays impact tcp performance for a connectivity from reunion island?'' in
  \emph{Next Generation Computing Applications (NextComp), 2017 1st
  International Conference on}.\hskip 1em plus 0.5em minus 0.4em\relax IEEE, 07
  2017, p. 103–108. [Online]. Available:
  \url{https://doi.org/10.1109/NEXTCOMP.2017.8016183}
\BIBentrySTDinterwordspacing

\bibitem{cablesmap}
Telegeography, ``Submarine cable map.''

\bibitem{axis}
``African internet exchange system project,'' 2016.

\bibitem{gmaps}
Google, ``Google maps,'' 2017.

\bibitem{Snyder2012}
A.~W. Snyder and J.~Love, \emph{Optical waveguide theory}.\hskip 1em plus 0.5em
  minus 0.4em\relax Springer Science \& Business Media, 2012.

\bibitem{Krajsa2011}
O.~Krajsa and L.~Fojtova, ``{RTT} measurement and its dependence on the real
  geographical distance,'' in \emph{Telecommunications and Signal Processing
  (TSP), 2011 34th International Conference on}.\hskip 1em plus 0.5em minus
  0.4em\relax IEEE, 2011, pp. 231--234.

\bibitem{Soobron2014}
M.~Soobron, C.~Soobron, S.~Soobron, A.~Sukhoo, and R.~H. Hawabhay,
  ``Connectivity within indian ocean islands (mauritius, seychelles, comoros,
  reunion and madagascar)—a case for a regional internet exchange,'' in
  \emph{IST-Africa Conference Proceedings, 2014}.\hskip 1em plus 0.5em minus
  0.4em\relax IEEE, 2014, pp. 1--18.

\bibitem{Galperin2013}
H.~Galperin, ``Connectivity in latin america and the caribbean: The role of
  internet exchange points,'' \emph{Internet Society, November}, 2013.

\end{thebibliography}
\bibliographystyle{IEEEtran}

%
%
%
%
%
%
%
%
%
%
\end{document}